# RA: A machine based rational agent
# Part 2
# Preliminary test

G. Pantelis[1]
May 26, 2024

**Abstract**

*A preliminary test of the software package RA is presented. The main focus of this test is to assess RA's reasoning capabilities that are based on the formal system PECR. Particular attention is given to the finite computational resources of the real-world machine that define the environment within which programs are to be executed.*

## 1 Introduction

RA is a software package that was designed to acquire an understanding of the laws that govern real-world phenomena working under the hypothesis of a computable universe. It perceives the fundamental properties of objects and their dynamic behavior through a language of computation expressed in the form of programs as outlined in [1] and [2].

An understanding of the dynamics of real-world phenomena is acquired through a process of creating models. RA's primary objective is to find the laws of real-world phenomena from which computer models can be constructed. RA regards a computer model to be valid if (i) it can be rigorously shown to be computable and (ii) it produces results that are consistent with observation. To address computability we must take into account the finite computational resources of the machine on which the computer model is to be executed. Therefore it is important to take some care in placing too much reliance on any theory of computation developed in the abstract realm of mathematics.

We write

M[mach]

to refer to a real-world computer as a machine, M, where mach is a list of parameters that constrain certain operations of programs so that their execution do not make excessive demands on the finite computational resources of the machine (Section 1.4 of [2]).

RA explores applications by executing the three actions given in the following table.

*Table 1: Actions of an application*

| Action | Description |
|---|---|
| 1 | Conjecture |
| 2 | Soundness checking |
| 3 | Theorem proving |

1 garrypantelis@proton.me

The overall process is one of iteration of ongoing revision and feedback of the parallel Actions 1-3. The process is initiated by finding conjectures that then undergo checks for soundness. Of the surviving conjectures proofs are sought.

Before RA explores any real-world application we test its main components on a simple example. Here we will present the results generated by RA for an exploration of the rules of arithmetic on the finite universe of objects of type nat. Objects of type nat are natural numbers that are bounded below by 0 and bounded above by a maximum machine number, mnat. They reflect the constraints that exist when performing operations of arithmetic over the natural numbers on a real-world computer.

In mathematics the natural numbers are defined as an infinite set, $\mathbb{N}$, and the operations of arithmetic on the natural numbers are governed by the Peano axioms. RA’s reasoning is based on the formal system PECR that was designed to address real-world computations. Because we are interested in a finite universe of objects of type nat we can expect that the rules of arithmetic in this universe will differ from those of Peano arithmetic. It is useful to remind ourselves of the rules of Peano arithmetic. For reference purposes they are presented in Appendix A1.

## 2 Program structure

In the formal system PECR we deal exclusively with program lists of elementary functional programs where each elementary functional program is referred to as an *atomic program* (AP). When representing APs we employ labels from the three lists

```
pname=[pname[1] pname[2] ... pname[nap]]
var=[var[1] var[2] ... var[nvar]]
cst=[cst[1] cst[2] ... cst[ncst]]
```

The elements of the list pname are the names of the APs that are specific to the application under consideration. Each element of the input list of an AP is a label taken from elements of var if it is a variable or elements of cst if it is a constant. Labels of elements of the output list of an AP are taken from elements of var only.

A program, p, of list length n can be represented as an ordered list

```
p=[p[1] p[2] ... p[n]]
```

Each p[i], i=1,2,...,n, is an AP that has the list structure

```
p[i]=[pn[i] x[i] y[i]]
```

where pn[i] is the program name of p[i] taken from the list pname, x[i] is the input list and y[i] is the output list. In the interests of a closer resemblance to actual code we sometimes represent a program as a vertical list of APs.

Underlying all applications are the APs from which the fundamental rules of the PECR language itself are expressed. These APs are *higher order programs* (HOPs). HOPs are functional programs whose I/O list elements can be assigned values of programs.

## 3 Atomic programs of arithmetic

In the formal system PECR objects are identified by their type and expressed in the form of scalars, lists and arrays. We denote the application of arithmetic on the finite universe of type nat objects by

nat[pname mach]

where pname is the list of names of the APs of the application and mach is a list of machine parameter constraints specific to the machine M[mach] upon which the application is to be explored.

The list of constants, cst, of the application nat[pname mach] is given by

cst=[0 1 mnat]

where mnat is the maximum machine number. Here the elements 0 and 1 should be regarded as alphanumeric character strings that are assigned the fixed values of zero and one, respectively.

All APs internally perform type checking of the value assignments of their I/O list elements. If there is a type violation the program halts with an error message. Type and relation checking programs have empty output lists and are given by the following table. The first column gives the integer label of the program name of the AP. The integer labels of program names will be employed later.

*Table 2: Type and relation checking atomic programs*

| **Integer label** | **Alphanumeric representation** | **Type and relation checks** |
|---|---|---|
| 1 | [typen [a] []] | type[a nat] |
| 2 | [eqn [a b] []] | type[a nat]<br>type[b nat]<br>eqn[a b] |
| 3 | [lt [a b] []] | type[a nat]<br>type[b nat]<br>lt[a b] |

The APs [eqn [a b] []] and [lt [a b] []] check for type violations as well as a relation condition. These programs will halt with an error message if there is a violation in type or relation. In conventional notation the terms eqn[a b] and lt[a b] are given by a=b and a<b, respectively.

The APs that perform the computations of addition and multiplication are given by Table 3.

*Table 3: Atomic programs of addition and multiplication*

| **Integer label** | **Alphanumeric representation** | **Type checks** | **Value and type assignment** |
|---|---|---|---|
| 4 | [add [a b] [c]] | type[a nat]<br>type[b nat] | c:=add[a b]<br>type[c nat] |
| 5 | [mult [a b] [c] | type[a nat]<br>type[b nat] | c:=mult[a b]<br>type[c nat] |

The APs [add [a b] [c]] and [mult [a b] [c]] will halt with an error message if there is a type violation of any value assignment of the elements of their I/O lists. In conventional notation the terms add[a b] and mult[a b] are given by a+b and a∗b, respectively.

## 4 Disjunctions

We introduce the special disjunction programs that are associated with the terms le[a b] and trich[a b]. The APs that are the realization of these terms are given in the following table.

*Table 4: Disjunctions*

| **Integer label** | **Alphanumeric representation** | **Type and relation checks** |
|---|---|---|
| 6 | [le [a b] []] | type[a nat]<br>type[b nat]<br>le[a b] |
| 7 | [trich [a b] []] | type[a nat]<br>type[b nat]<br>trich[a b] |

<u>Less than or equal</u>

The program [le [a b] []] is an AP that is constructed by the HOP

[disj [p q] [s]]

where p, q and s have the value assignments

val[p]=[lt [a b] []]
val[q]=[eqn [a b] []]
val[s]=[le [a b] []]

The following rules apply to the disjunction program [le [a b] []].

```
le 1                        le 2

lt [a b] []                 eqn [a b] []
-----------                 ------------
le [a b] []                 le [a b] []
```

<u>Trichotomy</u>

Having constructed the disjunction [le [a b] []] we can now introduce the trichotomy program that is constructed by the HOP

[disj [p q] [s]]

where p, q and s have the value assignments

val[p]=[lt [b a] []]
val[q]=[le [a b] []]
val[s]=[trich [a b] []]

Following the general structural rules of application specific disjunctions as outlined in Section 7.3 of [2] we could write

```
lt [b a] []                le [a b] []
--------------             --------------
trich [a b] []             trich [a b] []
```

However these are not axioms because they immediately follow from Ord 8 (Section 5) by applying the IOT axiom to both input variables, a and b.

## 5 Rules of arithmetic over nat

The difficulty with arithmetic on a machine M[mach] is the absence of closure for addition and multiplication. Because of this many of the rules of arithmetic on a machine M[mach] will be quite different to those of Peano arithmetic.

Before we test RA's ability to acquire an understanding of the rules of arithmetic in a finite universe of type nat objects we will attempt to put forward our own collection of rules. Some of these rules will come in pairs and triplets with one or more existence rule followed by an equality rule. We are cautious to make any claims whether any of these rules are axioms or theorems.

In PECR rules are expressed in the form of *irreducible extended programs* (IEPs) as outlined in Section 4.2 of [2]. An IEP takes the form

```
s=conc[p c]
type[c iext[p]]
```

and asserts that if the premise program, p, is computable for some value assignment of the elements of its primary input list then the concatenation, s=conc[p c], of the premise program, p, and the conclusion program, c, is also computable for the same value assignments. Crucial to all program lists in PECR is the adherence to the I/O dependency conditions (Section 3.2 of [2]).

In our constructive approach to computability we do not make use of quantifiers. We start by stating the following equality rules.

Equality is reflexive, symmetric and transitive

```
Nat 1a                     Nat 1b                     Nat 1c

typen [a] []               eqn [a b] []               eqn [a b] []
------------               ------------               eqn [b c] []
eqn [a a] []               eqn [b a] []               ------------
                                                      eqn [a c] []
```

These are analogous to Peano axioms eq_1, eq_2 and eq_3 (Appendix A1). Any attempt to state an analagy of the Peano axiom eq_4 in PECR would be rendered superfluous by the IOT axiom.

For the operations of addition and multiplication we have the following rules.

Addition is commutative

Nat 2a

```
add [a b] [c]
-------------
add [b a] [d]
```

Nat 2b

```
add [a b] [c]
add [b a] [d]
-------------
eqn [d c] []
```

Addition is associative

Nat 3a

```
add [a b] [d]
add [d c] [x]
add [b c] [e]
-------------
add [a e] [y]
```

Nat 3b

```
add [b c] [e]
add [a e] [y]
add [a b] [d]
-------------
add [d c] [x]
```

Nat 3c

```
add [a b] [d]
add [d c] [x]
add [b c] [e]
add [a e] [y]
-------------
eqn [y x] []
```

Addition with zero

Nat 4a

```
typen [a] []
-------------
add [0 a] [b]
```

Nat 4b

```
add [0 a] [b]
-------------
eqn [b a] []
```

Multiplication is commutative

Nat 5a

```
mult [a b] [c]
--------------
mult [b a] [d]
```

Nat 5b

```
mult [a b] [c]
mult [b a] [d]
--------------
eqn [d c] []
```

Multiplication with zero

Nat 6a

```
typen [a] []
--------------
mult [0 a] [b]
```

Nat 6b

```
mult [0 a] [b]
--------------
eqn [b a] []
```

Multiplication with one

Nat 7a

```
typen [a] []
--------------
mult [1 a] [b]
```

Nat 7b

```
mult [1 a] [b]
--------------
eqn [b a] []
```

Multiplication is associative

```
Nat 8a                   Nat 8b                   Nat 8c

mult [a b] [d]           mult [b c] [e]           mult [a b] [d]
mult [d c] [x]           mult [a e] [y]           mult [d c] [x]
mult [b c] [e]           mult [a b] [d]           mult [b c] [e]
--------------           --------------           mult [a e] [y]
mult [a e] [y]           mult [d c] [x]           --------------
                                                  eqn [y x] []
```

Multiplication distributes over addition

```
Nat 9a                   Nat 9b                   Nat 9c

add [b c] [d]            mult [a b] [u]           add [b c] [d]
mult [a d] [x]           mult [a c] [v]           mult [a d] [x]
mult [a b] [u]           add [u v] [y]            mult [a b] [u]
mult [a c] [v]           add [b c] [d]            mult [a c] [v]
--------------           --------------           add [u v] [y]
add [u v] [y]            mult [a d] [x]           --------------
                                                  eqn [y x] []
```

We include the following order rules.

Order is preserved under addition of the same element

Ord 1

```
lt [a b] []
add [a c] [x]
add [b c] [y]
-------------
lt [x y] []
```

Order is preserved under multiplication by the same positive element

Ord 2

```
lt [a b] []
lt [0 c] []
mult [a c] [x]
mult [b c] [y]
--------------
lt [x y] []
```

Inequality is transitive

Ord 3

```
lt [a b] []
lt [b c] []
-----------
lt [a c] []
```

Zero and one are distinct

```
Ord 4

-----------
lt [0 1] []
```

There is no machine number between 0 and 1

```
Ord 5

lt [0 a] []
-----------
le [1 a] []
```

The AP lt is irreflexive

```
Ord 6

lt [a a] []
-----------
false
```

Zero is the minimum machine number and mnat is the maximum machine number

```
Ord 7a                      Ord 7b

typen [a] []                typen [a] []
------------                --------------
le [0 a] []                 le [a mnat] []
```

Ordering satisfies trichotomy

```
Ord 8

typen [a] []
typen [b] []
--------------
trich [a b] []
```

The formal system PECR is based on computability logic that is implemented directly on a machine platform. Statements in PECR come in the form of an AP that can either be computable or uncomputable for a given value assignment of its input list. This differs from mathematics where statements come in the form of well defined formulas that are subject to the value assignments true or false. Hence when making comparisons between applications of PECR and theories in mathematics we can only speak in terms of analogies.

The above rules of arithmetic can be associated, by analogy, with the Peano axioms (Appendix A1):

- The Peano axiom 2 states that addition is commutative. On a machine M[mach] we need to split this into two parts, an existence rule followed by an equality rule. Given any valid value assignments of two variables a and b there is no guarantee that their sum will be type nat on a machine M[mach]. Nat 2a states that if the sum a+b is computable for some value assignments of a and b then the sum b+a is also computable for the same value assignments. The I/O dependency conditions include the requirement that all output list element labels of a program list must be distinct. Nat 2b equates the value assignments of the output variables of the two sums.

- Nat 3c is analogous to the Peano axiom 1. Nat 3a is an existence rule that states that if for some value assignments of a, b and c the sums a+b, (a+b)+c and b+c are computable then so is the sum a+(b+c). Nat 3b is an existence rule that states that if for some value assignments of a, b and c the sums b+c, a+(b+c) and a+b are computable then so is the sum (a+b)+c.

- Nat 4b is analogous to the Peano axiom 6 and follows the existence rule Nat 4a.

- Nat 5b is analogous to the Peano axiom 4 and follows the existence rule Nat 5a.

- Nat 6b is analogous to the Peano axiom 7 and follows the existence rule Nat 6a.

- Nat 7b is analogous to the Peano axiom 8 and follows the existence rule Nat 7a.

- Nat 8c is analogous to the Peano axiom 3 and follows two existence rules Nat 8a and Nat 8b.

- Nat 9c is analogous to the Peano axiom 5 and follows two existence rules Nat 9a and Nat 9b.

- Ord 1 and Ord 2 are analogous to the Peano axioms 12 and 13, respectively.

- Ord 3 is analogous to the Peano axiom 9.

- Ord 4 is analogous to the Peano axiom 15. The premise of Ord 4 is the empty program.

- Ord 5 is analogous to the Peano axiom 16.

- Ord 6 is analogous to the Peano axiom 10. It states that the program [lt [a a] []] is not computable for any value assignment of the elements of its input list.

- The Peano axiom 17 states that zero is the minimum number. Ord 7a is the analogy of this Peano axiom. However, we are working on a machine M[mach] so we need an additional rule, Ord 7b, as a statement for the maximum machine number.

- Finally the Peano axiom 11 is expressed in the form of Ord 8.

The above IEPs appear to be fundamental rules that can be stated as axioms. But this is not the case. We will see that RA will generate all of these rules as conjectures and will establish that some of them are theorems.

## 6 Integer representation of programs

Elements of the lists pname, var and cst can be represented by alphanumeric character strings. In Chapter 10 of [2] it was shown that labels of these lists could easily be represented by integers. The convention we adopt is to take each element u[i], i=1,2,...,nu, of a list

```
u=[u[1] u[2] ... u[nu]]
```

and assign to it the integer label

```
u[i]=i
```

In this way u becomes an ordered list of natural numbers. Each integer element of the list u can be translated back to an alphanumeric character string if desired.

A program p=[p[1] p[2] ... p[n]] is translated into an integer array as follows.

```
type[p prgm[n]]
subtype[prgm[n] arr[d]]
type[d vec[2]]
d=[n 1+nx+ny]
i=1,2,...,n
   type[pn[i] nat]
   type[x[i] vec[nx]]
   type[y[i] vec[ny]]
   type[p[i] vec[1+nx+ny]]
   p[i]=chain[pn[i] x[i] y[i]]
```

where arrays whose elements are type nat are assigned the type arr. The parameter dependent type arr[d] denotes type array of dimension list d. An array type arr of rank 1 is a vector, type vec. The rank of an array is the length of its dimension list. The scalar pn[i] is treated as a singleton (list) under the action of chain[:]. The action of chain[:] removes the outermost brackets, [], enclosing the lists it is acting on (Section 2.4 of [2]).

Some APs will have I/O lists that have shorter lengths than the maximum list lengths nx and ny. We maintain x[i] and y[i] are type vec[nx] and vec[ny], respectively, and define an empty element of these lists as a *null variable*.

Each element, p[i], i=1,2,...,n, of the program list p is an AP whose I/O lists, x[i] and y[i], have the expanded form

```
x[i]=[x[i][1] x[i][2] ... x[i][nx]]
y[i]=[y[i][1] y[i][2] ... y[i][ny]]
```

Elements of the I/O lists x[i] and y[i] are assigned the integer labels under the following convention.

```
x[i][j] = {k,      le[1 k] le[k nvar], x[i][j] is a variable associated with var[k]
          {nvar+m, le[1 m] le[m ncst], x[i][j] is a constant associated with cst[m]
          {0,      x[i][j] is a null variable

y[i][j] = {k,      le[1 k] le[k nvar], y[i][j] is a variable associated with var[k]
          {0,      y[i][j] is a null variable
```

Here we are effectively appending the list of constants, cst, to the list of variables, var. In other words an integer label nvar+m, le[1 m] le[m ncst], is associated with a constant cst[m] and an integer label k, le[1 k] le[k nvar], is associated with a variable var[k]. The length, nvar, of the list var is automatically set by RA.

Using this convention we can now write a program p, type[p prgm[n]], as an integer matrix

```
[pn[1] x[1][1] x[1][2] ... x[1][nx] y[1][1] y[1][2] ... y[1][ny]]
[pn[2] x[2][1] x[2][2] ... x[2][nx] y[2][1] y[2][2] ... y[2][ny]]
.                                                                .
.                                                                .
.                                                                .
[pn[n] x[n][1] x[n][2] ... x[n][nx] y[n][1] y[n][2] ... y[n][ny]]
```

The *I/0 matrix* of a program is obtained by removing the first column of this matrix. Crucial to the validity of any program list are the I/O dependency conditions (Section 3.2 of [2]).

Note that the integers in the I/O matrix are labels of the elements of the I/O lists and not value assignments of the I/O list elements. Value assignments require an additional map that sends each integer I/O element label to an object that has a type consistent with that checked in the AP.

RA perceives programs as integer arrays and the process of program constructions as maps on integer arrays (Chapter 10 of [2]). Programs are translated to and from integer and alphanumeric formats only for the convenience of humans who prefer to view certain objects of RA’s I/O data files in a symbolic form.

## 7 I/O binding matrices

We can identify structural similarities in the rules presented in the Section 5. They come in the form of I/O binding matrices. A convenient way to view these structures is to adopt the integer array format of programs as outlined in Section 6. For example the IEPs Nat 2a and 2b can be expressed as the integer matrices

```
Nat 2a                        Nat 2b

[4 1 2 3]                     [4 1 2 3]
[4 2 1 4]                     [4 2 1 4]
                              [2 4 3 0]
```

where we are using nx=2 and ny=1. The first columns of these matrices are the integer labels of the program names as presented in the first columns of Tables 2-4. Removing the first columns yield the I/O matrices. The integer variable labels are arbitrary provided they belong to the finite list of integer variable labels, var, and retain the same configurations of repetitions. Null variables are set to zero.

It is important to keep in mind that the integers in the I/O matrices are just labels of I/O list elements and are not value assignment of I/O list elements. In the application nat[pname mach] value assignments of elements of the I/O lists of programs are made through a separate map that sends each integer I/O list element label to an object of type nat.

Following the procedures outlined in Section 10.3 of [2] we decompose the I/O matrices to isolate each distinct label. Each matrix in such a decomposition is said to be an *I/O binding matrix* if it has more that one nonzero entry associated with a variable or a nonzero entry associated with a constant. We adopt the convention that an integer label of an element cst[m] of the list of constants cst is given by nvar+m, where nvar is the length of the list of variable integer labels var. In the above matrices all I/O elements are associated with variables.

For Nat 2a the I/O matrix can be decomposed into the matrices

```
[1 0 0]          [0 2 0]          [0 0 3]          [0 0 0]
[0 1 0]          [2 0 0]          [0 0 0]          [0 0 4]
```

The original I/O matrix is obtained by applying a standard matrix sum to these four matrices. The first two matrices represent I/O binding matrices while the last two matrices are not I/O binding matrices because they have no repetitions of variable labels and no labels bound to a constant.

Isolating each variable label we obtain for Nat 2b

```
[1 0 0]          [0 2 0]          [0 0 3]          [0 0 0]
[0 1 0]          [2 0 0]          [0 0 0]          [0 0 4]
[0 0 0]          [0 0 0]          [0 3 0]          [4 0 0]
```

All four of these matrices represent I/O binding matrices.

When comparing I/O matrices we are not interested in the actual numerical value of the integer variable labels but rather in the structural patterns that emerge from I/O element labels that are bound to each other or bound to a constant. It is easier to identify these patterns by using the *binary I/O matrix templates* of the above matrices (Section 10.3 of [2]). Binary I/O matrix templates are obtained by simply replacing all nonzero elements of the above matrices with 1. Each binary I/O matrix template must retain a connection to the original integer I/O element label to distinguish matrices associated with variables from matrices associated with constants. Two programs that have identical binary I/O binding matrix templates are said to be structurally equivalent.

Since we are interested in comparing binary I/O binding matrix templates we can discard the last two matrices associated with Nat 2a that are not binding matrices. We find that Nat 2a and 2b are structurally equivalent to the IEPs Nat 5a and 5b, respectively. We also find that Nat 2a is structural equivalent to Nat 1b.

In a similar way we find that the decomposition of the I/O matrices of the IEPs Nat 1c and Ord 3 are structurally equivalent.

So far we have examined IEPs with the simple structures of symmetry and transitivity. It would be convenient if similarities exist for IEPs of greater complexity. Indeed we find that the I/O matrix decompositions of IEPs Nat 3a, 3b and 3c are structurally equivalent to those of IEPs Nat 8a, 8b and 8c, respectively.

For examples of IEPs containing constants we find that Nat 4a, Nat 6a and Ord 7a are structurally equivalent while Nat 4b and Nat 6b are structurally equivalent.

There is no known formal procedure that can describe the process by which a human mind constructs conjectures. It appears to be some kind of intuitive process that is still not understood. When we view configuration states of I/O matrices associated with IEPs through I/O matrix decompositions we find that many IEPs have similar structures. This suggests that recognizing patterns that emerge from the I/O binding matrices may play an important role on how a rational agent, human or otherwise, identifies conjectures.

Adding further weight to this we find that many structural patterns of I/O matrices are not confined to a specific application. For example the same structural patterns of the I/O matrix decompositions of reflexivity, symmetry and transitivity can be observed across applications. In addition these structural similarities are type independent.

It is important to note that the I/O lists of any program can be completely determined by the binary I/O binding matrix templates alone. Consequently, RA need only explore configurations of binary I/O binding matrix templates. This combined with an elimination procedure based on the *program extension* (PE) structural integrity tests (Section 4.2 of [2]) significantly reduces the search for general structures of programs posed as conjectures. Embedded within the PE structural integrity tests are the I/O dependency conditions that must be satisfied by all programs.

The construction of conjectures through binary I/O binding matrix templates can in principle be fully automated. However, even under the filtering process through the structural PE integrity tests, an ordered and systematic search over all possible configuration states of binary I/O binding matrix templates for programs that have premises of length greater than three can be computationally expensive. There is a need to include additional filters that reduce the search for binary I/O binding matrix templates to those that are relevant to the specific application being considered. Thus the PE structural integrity test should be combined with checks for consistency with empirical data.

It should be mentioned that the notion of structural equivalence as defined above has meaning in a stronger sense when comparing programs whose AP program names follow the same configuration of repetitions in the first column of the integer matrix representation of the program. For each group of binary I/O binding matrix templates RA generates a collection of provisional conjectures by matching them with a vertical column list of all permutations of the labels of the AP names. Some of these turn out to be valid conjectures while lacking the structural similarities in the stronger sense mentioned above.

## 8 Test results

RA was run on a laptop :: Memory:16 GB RAM, Processor:11th Gen Intel® Core™ i7-1195G7@2.90GHz×8, OS:Linux. On this machine Actions 1-3 were actually executed as an ongoing sequential loop iteration rather than in parallel as originally described. Each iteration is initiated by the generation of a collection of conjectures from a distinct group of binary I/O binding matrix templates. All conjectures are subjected to ongoing checks for soundness. Surviving conjectures are fed into RA’s automated theorem prover where proofs are sought.

Constructing conjectures is the end product of checks for PE structural integrity and checks for soundness. However, there are conjectures that can survive these tests. They come in the form of *extended programs* (EPs) that are not IEPs (Section 4.2 of [2]). These can be eliminated by RA’s theorem prover that has the ability to reject and remove conjectures whose premises contain superfluous statements. It does this by the process of connection list reduction as outlined in Section 10.4 of [2].

The combined actions of conjecturing, soundness checking and theorem proving is one of iteration involving ongoing revision and feedback. As part of this iteration RA continually updated the partitioned list of known IEPs

s=[ax th ud]

where ax is the list of axioms, th is the list of theorems and ud is the list of IEPs for which no proof is known.

To avoid a comprehensive and lengthy exploration of the application nat[name mach] artificial constraints were placed on the number of binary I/O binding matrix templates that RA could access. RA halted when convergence was established, i.e. when all proofs and the IEP partition list, s, remained unchanged. The final output is presented in Appendix A2 for reference purposes. Under the artificial constraints imposed on the conjecturing process and the machine resource constraints set by the machine parameters list, mach, convergence was reached in just under 1 minute (wall clock). In this time RA found 115 IEPs that survived the elimination procedures contained in each of Actions 1-3. Of these RA identified 28 as axioms and found proofs for 79. The remainder 8 were designated as underivable.

Underivable IEPs are defined as such only with respect to the machine parameter constraints mach and the known IEPs of the sub-application under consideration. It is worth noting that some of these were identified as axioms and theorems in another expanded application of nat[pname mach], the results of which are not presented here for brevity.

Axioms and theorems along with their proofs are clearly labeled. Each axiom and theorem is followed by the IEP integer label. The bracketed lists following the theorem labels are the theorem connection lists that contain the integer labels of the IEPs that were explicitly employed in the proof of the theorem. Theorem connection lists exclude the IOT axiom and the substitution rule. Underivables are just labeled by the string UD followed by an IEP integer label.

On the far right of each axiom/theorem label there is a number that quantifies the weight of the axiom/theorem. The weight of an IEP is the number of times it can be traced back as a dependency in proofs of theorems.

Each derived statement that appears in a proof is an AP that is preceded by an integer statement label. On the far right of each derived statement is the integer label of the IEP that was employed in the derivation of that statement followed by a sublist connection list (not to be confused with the theorem connection list). The sublist connection list contains the integer labels of the statements (APs) that formed the sublist of the proof program that was I/O equivalent to the premise of the IEP whose label precedes the sublist connection list. A statement that is derived by the splitting of a disjunction is followed by two pairs of IEP label and sublist connection list, each pair of IEP label and sublist connection list being associated with an operand of the disjunction. The IEP label dcr2 means that the *disjunction contraction rule 2* has been applied to that operand of the disjunction (Section 6.3 of [2]).

## 9 Analysis of test results

The rules of arithmetic on a real-world machine, M[mach], require a major shift in mindset from one that was nurtured under the ideal world of Peano arithmetic. Consider the two programs [add [a b] [c]] and [mult [a b] [c]]. Given any valid value assignments of the input elements a and b there is no guarantee that either of these programs will be computable on a machine M[mach]. In Peano arithmetic this is not an issue because ℕ is closed under addition and multiplication.

A somewhat trivial result in the form of theorem 60 (Appendix A2) is a striking example that addresses this issue, reproduced here for convenience.

Theorem 60

```
add [a b] [c]
add [d c] [e]
-------------
add [d b] [j]
```

Theorem 60 asserts that if for some value assignments of a and b the sum of a and b exists and if for some value assignment of d the sum of d and a+b exists then so does the sum of d and b. In Peano arithmetic this rule is superfluous but it can be an important rule to access when a rational agent is attempting to establish, by some automated process, the computability of a program on a real-world computer. Indeed theorem 60 is employed in the proof of theorem 74 of which more will be said shortly.

We see that all of the human constructed rules of Section 5 are contained in the final collection of conjectures generated by RA as shown in Appendix A2, although not in the order as they appear in Section 5. RA established that some of the human constructed rules of Section 5 are theorems. For example reflexivity and symmetry of equality, labeled Nat 1a and Nat 1b, respectively, are presented as theorems 2 and 10, respectively. Transitivity of equality, Nat 1c, is not shown because it follows from the substitution rule of PECR.

The associative rule of addition, Nat 3c, is preceded by two existence rules, Nat 3a and 3b. Theorem 74 demonstrates that the second existence rule, Nat 3b, follows from the first existence rule Nat 3a (axiom 66).

Similarly, the associative rule of multiplication, Nat 8c, is preceded by two existence rules, Nat 8a and 8b. Theorem 76 demonstrates that the second existence rule, Nat 8b, follows from the first existence rule Nat 8a (axiom 69).

The following theorem is also worth noting.

Theorem 92

```
add [a b] [c]
add [d b] [e]
eqn [e c] []
-------------
eqn [d a] []
```

Theorem 92 asserts that if for some value assignments of a, b, and d the sums a+b and d+b exist and are equal then it follows that d and a are equal. In the application nat[pname mach] we do not have an operation of subtraction so this result cannot be derived using basic high school algebra. The proof of this theorem employs disjunction splitting that depends on a theorem of falsity, theorem 91.

Similarly, consider the following theorem.

Theorem 111

```
lt [0 a] []
mult [b a] [c]
mult [d a] [e]
eqn [e c] []
--------------
eqn [d b] []
```

Theorem 111 asserts that if for some value assignments of a, b and d the products b*a and d*a exist and are equal and a is greater than zero then d and b are equal. In the application nat[pname mach] we do not have an operation of division so again this result cannot be derived using basic high school algebra. The proof of this theorem employs disjunction splitting that depends on a theorem of falsity, theorem 102.

In [1] it was mentioned that RA’s automated theorem prover can be employed as a stand alone package. Most of the proofs of theorems that were generated by RA for various applications are very similar to those that were originally constructed interactively with VPC. However, there are instances where RA exhibits strategies that appear to be alien to human constructed proofs.

Examples of this can be found here in the form of theorems 35, 36, 38 and 40. Their proofs simply apply the IOT axiom to a variable in the I/O lists of the premise program from which a conclusion is obtained. While each of these are valid IEPs their premises appear to a human as having little meaningful connection to their conclusions. Theorems 22, 23, 25 and 26 are similar but require an additional step employing a symmetry rule.

A human would likely discard these theorems as being too trivial and lacking in informative value to be stored as IEPs. However, consider the IEP 31

```
-------------
add [0 1] [c]
```

The conclusion of IEP 31 is just a specific case of the conclusion of IEP 4. However IEP 31 has an empty premise and as such RA correctly identified it as an independent conjecture. RA then went on to find a proof of IEP 31 employing IEPs 35 and 36, the latter two IEPS are mentioned above as likely to have been discarded by a human.

There are other theorems generated by RA that also have proofs that appear to be atypical to human constructed proofs. To understand this phenomena we must look at the application in its entirety. RA attempts to find as many theorems as possible that have proofs that are completely independent in the sense discussed in Section 5 of [1]. The process involves eliminating proofs and searching for alternative proofs until this is achieved. In this way proofs that make more sense to a human can be discarded to make way for less obvious proofs that are otherwise independent.

One of the main objectives in the construction of the software package RA is to employ well defined primitives and avoid strategies that are based on sophisticated algorithms. While such strategies may be powerful from the perspective of computational efficiency they are often of a heuristic nature that can sometimes lack robustness. The primitive algorithms that we are interested in should possess the property of robustness and, while sometimes being more computational expensive, have a global reach across all applications.

To get some idea of what these primitive algorithms might or might not look like it is sometimes useful to turn to nature for guidance. For example, current neuron models employed in neural networks are based on the evaluation of a weighted sum of the input signals. The weights are obtained from an optimization process that employs the gradient descent method. While regarded as an elegant and powerful tool to those that are more mathematically inclined, the gradient descent method is, to some extent, a human construct that is unlikely to have emerged through evolution by natural processes. There are effective neural network models that are algorithmically primitive in structure and avoid the use of weighted sums. This topic will be discussed in more detail elsewhere.

It is difficult to avoid some heuristics creeping into the process whereby RA partitions the list of conjectures into axioms, theorems and underivables. The choice is based on a strategy that does not necessarily lead to an optimal partition. For example the transitivity of inequality, expressed in PECR by IEP 41, is usually regarded as an axiom. Here RA found proofs of the IEPs 41 and 48 that were not independent of each other and has chosen to present IEP 41 as a theorem and regarded IEP 48 as an axiom. IEP 48 has a weight 31 that is greater than the weight 27 of IEP 41 suggesting, although not strongly, that the choice might have been an appropriate one.

This is just another component of RA that is marked for further investigation. There is a wide range of alternative strategies for this particular feature of RA that can be explored in future versions.

It was found that run times were significantly effected by the order in which conjectures were generated. RA continuously seeks proofs of IEPs with respect to all of the other known IEPs. As the number of IEPs of an application grows through RA’s continual conjecturing process the demand on computational resources increases. At some point this demand on computational resources can impair RA’s ability to find proofs. This impairment depends, to some extent, on the size of the machine on which RA is implemented.

On these observations it was decided to break up the application nat[pname mach] into several sub-applications. A sub-application generates conjectures from a restricted collection of binary I/O binding matrix template configurations. As a result the iteration of revision and feedback under Actions 1-3 led to a faster convergence of the partition of the list s=[ax th ud].

It is in this aspect of the application considered here where human intervention was the most intrusive. A collection of human defined restrictions were applied on the configurations of I/O binding matrices that RA could access from which it could generate conjectures. A major difficulty here is that setting artificial constraints on the conjecturing process can be subject to bias that is motivated by many factors. For humans these factors can sometimes fall outside of the realm of any well defined formal procedures making it difficult to automate.

The test example considered here was chosen for its simplicity and familiarity to humans as a topic while possessing properties that would allow for the assessment of RA’s main reasoning capabilities. However as an application, arithmetic on the natural numbers is a construct that has more of an appeal to mathematicians and differs in some ways from the task of searching for natural laws through the observation of real-world phenomena. There are good reasons to expect that many of the artificial intrusions mentioned above, especially on the conjecturing process, will not be necessary when RA is applied to real-world applications.

In Section 1 of [1] we alluded to the fact that RA has no sensory apparatus to directly perceive the real-world. Therefore there is no other option in the foreseeable future for RA to rely on the supply of real-world data by an external human agent. Putting this aside there still remains additional data that needs to be supplied to RA when setting up each application. This includes the application specific APs that are defined in advance by a human. The current focus of RA's development is to remove all of these external dependencies by endowing RA with the ability to identify all of the data required to initialize a specific application.

To address the removal of these artificial external intrusions we need to focus on real-world applications that was the original motivation behind the construction of RA. Fortunately we have some understanding of the general structure and properties of the main core of computer models of real-world phenomena. These are outlined in Sections 9.7 and 9.8 of [2] and come in the form of iterations of maps over arrays. The main task is to investigate the properties of these maps that will autonomously adapt to each application specific empirical dataset. A deeper exploration into this topic is highly influenced by the ideas discussed in Section 1.4 of [2].

## References

[1] G. Pantelis, *RA: A Machine Based Rational Agent, Part 1*, ArXiv:2405.12551,2024.
[2] G. Pantelis, *PECR: A formal system based on computability logic*, ArXiv:2403.14880, 2024.

## Appendix A1

**Peano axioms**

The Peano axioms apply to the natural numbers represented as a set denoted by $\mathbb{N}$. They start by introducing two constant symbols 0 and 1 and the successor function along with the equality axioms ::

eq_1. For every natural number x, x=x, i.e. equality is reflexive.

eq_2. For all natural numbers x and y, if x=y then y=x, i.e. equality is symmetric.

eq_3. For all natural numbers x, y and z, if x=y and y=z then x=z, i.e. equality is transitive.

eq_4. For all x and y, if y is a natural number and x=y, then x is also a natural number, i.e. the natural numbers are closed under equality.

Here we will use an equivalent axiomatization of the Peano axioms under first order logic. The rules for arithmetic on the natural numbers under this axiomatization are as follows. The above equality axioms are not included since they are logically valid in first-order logic with equality.

1. ∀x,y,z((x+y)+z=x+(y+z)), i.e. addition is associative.

2. ∀x,y(x+y=y+x), i.e. addition is commutative.

3. ∀x,y,y((x∗y)∗z=x∗(y∗z)), i.e. multiplication is associative.

4. ∀x,y(x∗y=y∗x), i.e. multiplication is commutative.

5. ∀x,y,z(x∗(y+z)=x∗y+x∗z), i.e. multiplication distributes over addition.

6. ∀x(x+0=x), i.e. zero is an identity for addition

7. ∀x(x∗0=0), i.e. zero is an absorbing element for multiplication.

8. ∀x(x∗1=x), i.e. one is an identity for multiplication.

9. ∀x,y,z(x<y∧y<z)→x<z), i.e. the '<' operator is transitive.

10. ∀x,y(¬(x<x)), i.e. the '<' operator is irreflexive.

11. ∀x,y(x<y∨x=y∨y<x), i.e. the ordering satisfies trichotomy.

12. ∀x,y,z(x<y→x+z<y+z), i.e. the ordering is preserved under addition of the same element.

13. ∀x,y,z(0<z∧x<y→x∗z<y∗z), i.e. the ordering is preserved under multiplication by the same positive element.

14. ∀x,y(x<y→∃z(x+z=y)), i.e. given any two distinct elements, the larger is the smaller plus another element.

15. 0<1, i.e. zero and one are distinct

16. ∀x(x>0→x≥1), i.e. there is no element between zero and one

17. ∀x(x≥0), i.e. zero is the minimum element.

We can omit axiom 7 because it can be derived from the other axioms.

## Appendix A2

**Final output of a sub-application of nat[pname mach]**

It is important to recall that in PECR an IEP should be read as follows :: If the premise program is computable for some value assignment of the elements of its primary input list then the concatenation of the premise program with the conclusion program is computable for the same value assignments. IEPs of falsity have a conclusion that simply states false. This means that the premise program is not computable for any value assignment of its primary input list.

```
Axiom  1                                                      45

typen [a] []
typen [b] []
--------------
trich [a b] []

Theorem  2  [3 9]                                             16

typen [a] []
------------
eqn [a a] []

proof

    1  typen [a] []
    2  le [a a] []                     3 [1]
    3  eqn [a a] []                    9 [2] dcr2 [2]
 eop

Theorem  3  [1 9]                                             27

typen [a] []
------------
le [a a] []
```

```
proof

    1  typen [a] []
    2  trich [a a] []                    1 [1 1]
    3  le [a a] []                       9 [2] dcr2 [2]
 eop

Theorem  4  [7 30 44]                                          5

typen [a] []
-------------
add [0 a] [c]

proof

    1  typen [a] []
    2  mult [1 a] [b]                    7 [1]
    3  lt [0 1] []                       30 []
    4  add [0 a] [c]                     44 [3 2]
 eop

Theorem  5  [7 30 45]                                          3

typen [a] []
--------------
mult [0 a] [c]

proof

    1  typen [a] []
    2  mult [1 a] [b]                    7 [1]
    3  lt [0 1] []                       30 []
    4  mult [0 a] [c]                    45 [3 2]
 eop

Theorem  6  [4 16 17]                                          1

typen [a] []
------------
le [0 a] []

proof

    1  typen [a] []
    2  add [0 a] [b]                     4 [1]
    3  le [0 b] []                       16 [2]
    4  eqn [b a] []                      17 [2]
    5  le [0 a] []                       sr1 [3 4]
 eop

Axiom  7                                                      17

typen [a] []
--------------
mult [1 a] [c]

UD  8                                                          0

typen [a] []
--------------
le [a mnat] []

Axiom  9                                                      49

lt [a a] []
-----------
false

Theorem  10  [2]                                              15

eqn [a b] []
```

```
------------
eqn [b a] []

proof

    1  eqn [a b] []
    2  typen [a] []                   iot [1]
    3  eqn [a a] []                   2 [2]
    4  eqn [b a] []                   sr1 [3 1]
 eop

Theorem  11  [1 9]                                           32

eqn [a b] []
------------
le [b a] []

proof

    1  eqn [a b] []
    2  typen [a] []                   iot [1]
    3  trich [a a] []                 1 [2 2]
    4  le [b a] []                    9 [3] sr1 [3 1]
 eop

Axiom  12                                                    20

add [a b] [c]
-------------
add [b a] [d]

Axiom  13                                                    12

mult [a b] [c]
--------------
mult [b a] [d]

Theorem  14  [10 11]                                          9

eqn [a b] []
------------
le [a b] []

proof

    1  eqn [a b] []
    2  eqn [b a] []                   10 [1]
    3  le [a b] []                    11 [2]
 eop

Axiom  15                                                    44

lt [a b] []
-----------
le [a b] []

Axiom  16                                                    22

add [a b] [c]
-------------
le [a c] []

Axiom  17                                                     4

add [0 a] [b]
-------------
eqn [b a] []

Theorem  18  [17 14]                                          0

add [0 a] [b]
```

```
-------------
le [b a] []

proof

    1  add [0 a] [b]
    2  eqn [b a] []                     17 [1]
    3  le [b a] []                      14 [2]
 eop

Theorem  19  [4 24 10]                                          0

mult [0 a] [b]
--------------
add [b a] [d]

proof

    1  mult [0 a] [b]
    2  typen [a] []                     iot [1]
    3  add [0 a] [c]                    4 [2]
    4  eqn [b 0] []                     24 [1]
    5  eqn [0 b] []                     10 [4]
    6  add [b a] [d]                    sr1 [3 5]
 eop

Theorem  20  [24 10]                                            0

mult [0 a] [b]
--------------
mult [b a] [c]

proof

    1  mult [0 a] [b]
    2  eqn [b 0] []                     24 [1]
    3  eqn [0 b] []                     10 [2]
    4  mult [b a] [c]                   sr1 [1 3]
 eop

Theorem  21  [6 24 10]                                          0

mult [0 a] [b]
--------------
le [b a] []

proof

    1  mult [0 a] [b]
    2  typen [a] []                     iot [1]
    3  le [0 a] []                      6 [2]
    4  eqn [b 0] []                     24 [1]
    5  eqn [0 b] []                     10 [4]
    6  le [b a] []                      sr1 [3 5]
 eop

Theorem  22  [4 12]                                             0

add [0 a] [b]
-------------
add [b 0] [d]

proof

    1  add [0 a] [b]
    2  typen [b] []                     iot [1]
    3  add [0 b] [c]                    4 [2]
    4  add [b 0] [d]                    12 [3]
 eop

Theorem  23  [5 13]                                             1
```

```
add [0 a] [b]
--------------
mult [b 0] [d]

proof

    1  add [0 a] [b]
    2  typen [b] []                    iot [1]
    3  mult [0 b] [c]                  5 [2]
    4  mult [b 0] [d]                  13 [3]
 eop

Axiom  24                                                  4

mult [0 a] [b]
--------------
eqn [b 0] []

Theorem  25  [4 12]                                        0

mult [0 a] [b]
--------------
add [b 0] [d]

proof

    1  mult [0 a] [b]
    2  typen [b] []                    iot [1]
    3  add [0 b] [c]                   4 [2]
    4  add [b 0] [d]                   12 [3]
 eop

Theorem  26  [5 13]                                        0

mult [0 a] [b]
--------------
mult [b 0] [d]

proof

    1  mult [0 a] [b]
    2  typen [b] []                    iot [1]
    3  mult [0 b] [c]                  5 [2]
    4  mult [b 0] [d]                  13 [3]
 eop

Theorem  27  [24 14]                                       0

mult [0 a] [b]
--------------
le [b 0] []

proof

    1  mult [0 a] [b]
    2  eqn [b 0] []                    24 [1]
    3  le [b 0] []                     14 [2]
 eop

Axiom  28                                                  1

mult [1 a] [b]
--------------
eqn [b a] []

Theorem  29  [28 14]                                       0

mult [1 a] [b]
--------------
le [b a] []
```

```
proof

    1  mult [1 a] [b]
    2  eqn [b a] []                    28 [1]
    3  le [b a] []                     14 [2]
 eop

Axiom  30                                                 13

-----------
lt [0 1] []

Theorem  31  [30 33 36 35 44]                              2

-------------
add [0 1] [b]

proof

    1  lt [0 1] []                     30 []
    2  le [0 1] []                     33 []
    3  mult [1 1] [a]                  36 [1] 35 [2]
    4  add [0 1] [b]                   44 [1 3]
 eop

Theorem  32  [31 17 23 13]                                 0

--------------
mult [0 1] [d]

proof

    1  add [0 1] [a]                   31 []
    2  eqn [a 1] []                    17 [1]
    3  mult [a 0] [b]                  23 [1]
    4  mult [1 0] [c]                  sr1 [3 2]
    5  mult [0 1] [d]                  13 [4]
 eop

Theorem  33  [30 15]                                       3

-----------
le [0 1] []

proof

    1  lt [0 1] []                     30 []
    2  le [0 1] []                     15 [1]
 eop

Theorem  34  [31 12]                                       0

eqn [0 a] []
-------------
add [1 a] [d]

proof

    1  eqn [0 a] []
    2  add [0 1] [b]                   31 []
    3  add [a 1] [c]                   sr1 [2 1]
    4  add [1 a] [d]                   12 [3]
 eop

Theorem  35  [7]                                           3

eqn [0 a] []
--------------
mult [1 a] [b]
```

```
proof

    1  eqn [0 a] []
    2  typen [a] []                      iot [1]
    3  mult [1 a] [b]                    7 [2]
 eop

Theorem  36  [7]                                               3

lt [0 a] []
--------------
mult [1 a] [b]

proof

    1  lt [0 a] []
    2  typen [a] []                      iot [1]
    3  mult [1 a] [b]                    7 [2]
 eop

UD  37                                                         0

lt [0 a] []
-----------
le [1 a] []

Theorem  38  [7]                                               0

add [0 a] [b]
--------------
mult [1 a] [c]

proof

    1  add [0 a] [b]
    2  typen [a] []                      iot [1]
    3  mult [1 a] [c]                    7 [2]
 eop

Theorem  39  [7]                                               0

mult [0 a] [b]
--------------
mult [1 a] [c]

proof

    1  mult [0 a] [b]
    2  typen [a] []                      iot [1]
    3  mult [1 a] [c]                    7 [2]
 eop

Theorem  40  [7]                                               0

le [0 a] []
--------------
mult [1 a] [b]

proof

    1  le [0 a] []
    2  typen [a] []                      iot [1]
    3  mult [1 a] [b]                    7 [2]
 eop

Theorem  41  [15 48]                                          27

lt [a b] []
lt [b c] []
-----------
lt [a c] []
```

```
proof

    1  lt [a b] []
    2  lt [b c] []
    3  le [a b] []                      15 [1]
    4  lt [a c] []                      48 [3 2]
 eop

Theorem  42  [41 15]                                     0

lt [a b] []
lt [b c] []
-----------
le [a c] []

proof

    1  lt [a b] []
    2  lt [b c] []
    3  lt [a c] []                      41 [1 2]
    4  le [a c] []                      15 [3]
 eop

Theorem  43  [15 50]                                     6

lt [a b] []
add [b c] [d]
-------------
add [a c] [e]

proof

    1  lt [a b] []
    2  add [b c] [d]
    3  le [a b] []                      15 [1]
    4  add [a c] [e]                    50 [3 2]
 eop

Axiom  44                                               12

lt [a b] []
mult [b c] [d]
--------------
add [a c] [f]

Theorem  45  [15 51]                                     4

lt [a b] []
mult [b c] [d]
--------------
mult [a c] [e]

proof

    1  lt [a b] []
    2  mult [b c] [d]
    3  le [a b] []                      15 [1]
    4  mult [a c] [e]                   51 [3 2]
 eop

Theorem  46  [41]                                       24

lt [a b] []
le [b c] []
-----------
lt [a c] []

proof

    1  lt [a b] []
```

```
   2  le [b c] []
   3  lt [a c] []                     41 [1 2] sr1 [1 2]
 eop

Theorem  47  [46 15]                                       0

lt [a b] []
le [b c] []
-----------
le [a c] []

proof

   1  lt [a b] []
   2  le [b c] []
   3  lt [a c] []                     46 [1 2]
   4  le [a c] []                     15 [3]
 eop

Axiom  48                                                 31

le [a b] []
lt [b c] []
-----------
lt [a c] []

Theorem  49  [48 15]                                      11

le [a b] []
lt [b c] []
-----------
le [a c] []

proof

   1  le [a b] []
   2  lt [b c] []
   3  lt [a c] []                     48 [1 2]
   4  le [a c] []                     15 [3]
 eop

Axiom  50                                                 12

le [a b] []
add [b c] [d]
-------------
add [a c] [f]

Axiom  51                                                 11

le [a b] []
mult [b c] [d]
--------------
mult [a c] [f]

Theorem  52  [49]                                         10

le [a b] []
le [b c] []
-----------
le [a c] []

proof

   1  le [a b] []
   2  le [b c] []
   3  le [a c] []                     49 [1 2] sr1 [1 2]
 eop

Theorem  53  [41 9]                                        0
```

```
lt [a b] []
lt [b a] []
-----------
false

proof

    1  lt [a b] []
    2  lt [b a] []
    3  lt [a a] []                     41 [1 2]
    4  false                           9 [3]
 eop

Theorem  54  [46 9]                                        16

lt [a b] []
le [b a] []
-----------
false

proof

    1  lt [a b] []
    2  le [b a] []
    3  lt [a a] []                     46 [1 2]
    4  false                           9 [3]
 eop

Axiom  55                                                  15

add [a b] [c]
add [b a] [d]
-------------
eqn [d c] []

Theorem  56  [55 11]                                       14

add [a b] [c]
add [b a] [d]
-------------
le [d c] []

proof

    1  add [a b] [c]
    2  add [b a] [d]
    3  eqn [c d] []                    55 [2 1]
    4  le [d c] []                     11 [3]
 eop

Axiom  57                                                   3

mult [a b] [c]
mult [b a] [d]
--------------
eqn [d c] []

Theorem  58  [57 11]                                        1

mult [a b] [c]
mult [b a] [d]
--------------
le [d c] []

proof

    1  mult [a b] [c]
    2  mult [b a] [d]
    3  eqn [c d] []                    57 [2 1]
    4  le [d c] []                     11 [3]
 eop
```

```
Theorem  59  [54]                                          6

le [a b] []
le [b a] []
------------
eqn [a b] []

proof

    1  le [a b] []
    2  le [b a] []
    3  eqn [a b] []                    54 [1 2] dcr2 [1]
 eop

Theorem  60  [12 16 56 50]                                 1

add [a b] [c]
add [d c] [e]
-------------
add [d b] [j]

proof

    1  add [a b] [c]
    2  add [d c] [e]
    3  add [b a] [f]                   12 [1]
    4  add [c d] [g]                   12 [2]
    5  le [b f] []                     16 [3]
    6  le [f c] []                     56 [1 3]
    7  add [f d] [h]                   50 [6 4]
    8  add [b d] [i]                   50 [5 7]
    9  add [d b] [j]                   12 [8]
 eop

Theorem  61  [12 13 16 56 51]                              0

add [a b] [c]
mult [d c] [e]
--------------
mult [d b] [j]

proof

    1  add [a b] [c]
    2  mult [d c] [e]
    3  add [b a] [f]                   12 [1]
    4  mult [c d] [g]                  13 [2]
    5  le [b f] []                     16 [3]
    6  le [f c] []                     56 [1 3]
    7  mult [f d] [h]                  51 [6 4]
    8  mult [b d] [i]                  51 [5 7]
    9  mult [d b] [j]                  13 [8]
 eop

Theorem  62  [3 72 46]                                     2

lt [a b] []
lt [c d] []
mult [b d] [e]
--------------
lt [a e] []

proof

    1  lt [a b] []
    2  lt [c d] []
    3  mult [b d] [e]
    4  typen [b] []                    iot [1]
    5  le [b b] []                     3 [4]
    6  le [b e] []                     72 [5 2 3]
```

```
    7  lt [a e] []                     46 [1 6]
 eop

Theorem  63  [62 15]                                        0

lt [a b] []
lt [c d] []
mult [b d] [e]
--------------
le [a e] []

proof

    1  lt [a b] []
    2  lt [c d] []
    3  mult [b d] [e]
    4  lt [a e] []                     62 [1 2 3]
    5  le [a e] []                     15 [4]
 eop

Theorem  64  [12 16 48 56 52 46]                            2

add [a b] [c]
lt [c d] []
add [b d] [e]
-------------
lt [a e] []

proof

    1  add [a b] [c]
    2  lt [c d] []
    3  add [b d] [e]
    4  add [d b] [f]                   12 [3]
    5  le [a c] []                     16 [1]
    6  le [d f] []                     16 [4]
    7  lt [a d] []                     48 [5 2]
    8  le [f e] []                     56 [3 4]
    9  le [d e] []                     52 [6 8]
   10  lt [a e] []                     46 [7 9]
 eop

Theorem  65  [64 15]                                        0

add [a b] [c]
lt [c d] []
add [b d] [e]
-------------
le [a e] []

proof

    1  add [a b] [c]
    2  lt [c d] []
    3  add [b d] [e]
    4  lt [a e] []                     64 [1 2 3]
    5  le [a e] []                     15 [4]
 eop

Axiom  66                                                   1

add [a b] [c]
add [c d] [e]
add [b d] [f]
-------------
add [a f] [m]

Axiom  67                                                   1

add [a b] [c]
mult [c d] [e]
```

```
mult [b d] [f]
--------------
add [a f] [m]

Theorem  68  [12 16 49 56 52]                               0

add [a b] [c]
le [c d] []
add [b d] [e]
-------------
le [a e] []

proof

    1  add [a b] [c]
    2  le [c d] []
    3  add [b d] [e]
    4  add [d b] [f]                 12 [3]
    5  le [a c] []                   16 [1]
    6  le [d f] []                   16 [4]
    7  le [a d] []                   49 [5 2] sr1 [5 2]
    8  le [f e] []                   56 [3 4]
    9  le [d e] []                   52 [6 8]
   10  le [a e] []                   52 [7 9]
 eop

Axiom  69                                                   1

mult [a b] [c]
mult [c d] [e]
mult [b d] [f]
--------------
mult [a f] [m]

Theorem  70  [12 43 56 84 16 46 48 52]                      5

le [a b] []
lt [c d] []
add [b d] [e]
-------------
lt [a e] []

proof

    1  le [a b] []
    2  lt [c d] []
    3  add [b d] [e]
    4  add [d b] [f]                 12 [3]
    5  add [c b] [g]                 43 [2 4]
    6  le [f e] []                   56 [3 4]
    7  add [b c] [h]                 12 [5]
    8  lt [g f] []                   84 [2 5 4]
    9  le [b h] []                   16 [7]
   10  lt [g e] []                   46 [8 6]
   11  le [h g] []                   56 [5 7]
   12  lt [h e] []                   48 [11 10]
   13  le [a h] []                   52 [1 9]
   14  lt [a e] []                   48 [13 12]
 eop

Theorem  71  [70 15]                                        0

le [a b] []
lt [c d] []
add [b d] [e]
-------------
le [a e] []

proof

    1  le [a b] []
```

```
    2  lt [c d] []
    3  add [b d] [e]
    4  lt [a e] []                      70 [1 2 3]
    5  le [a e] []                      15 [4]
 eop

Axiom  72                                                  9

le [a b] []
lt [c d] []
mult [b d] [e]
--------------
le [a e] []

Theorem  73  [16 54]                                       0

lt [a b] []
add [b c] [d]
eqn [a d] []
-------------
false

proof

    1  lt [a b] []
    2  add [b c] [d]
    3  eqn [a d] []
    4  le [b d] []                      16 [2]
    5  lt [d b] []                      sr1 [1 3]
    6  false                            54 [5 4]
 eop

Theorem  74  [12 60 66]                                    0

add [a b] [c]
add [b d] [e]
add [a e] [f]
-------------
add [c d] [m]

proof

    1  add [a b] [c]
    2  add [b d] [e]
    3  add [a e] [f]
    4  add [e a] [g]                    12 [3]
    5  add [a d] [h]                    60 [2 3]
    6  add [d a] [i]                    12 [5]
    7  add [b i] [j]                    66 [2 4 6]
    8  add [i b] [k]                    12 [7]
    9  add [d c] [l]                    66 [6 8 1]
   10  add [c d] [m]                    12 [9]
 eop

Theorem  75  [12 13 56 57 51 67]                           0

mult [a b] [c]
add [b d] [e]
mult [a e] [f]
--------------
add [c d] [m]

proof

    1  mult [a b] [c]
    2  add [b d] [e]
    3  mult [a e] [f]
    4  add [d b] [g]                    12 [2]
    5  mult [b a] [h]                   13 [1]
    6  mult [e a] [i]                   13 [3]
    7  le [g e] []                      56 [2 4]
```

```
    8  eqn [h c] []                    57 [1 5]
    9  mult [g a] [j]                  51 [7 6]
   10  add [d h] [k]                   67 [4 9 5]
   11  add [d c] [l]                   sr1 [10 8]
   12  add [c d] [m]                   12 [11]
 eop

Theorem  76  [13 57 58 51 69]                              0

mult [a b] [c]
mult [b d] [e]
mult [a e] [f]
--------------
mult [c d] [m]

proof

    1  mult [a b] [c]
    2  mult [b d] [e]
    3  mult [a e] [f]
    4  mult [b a] [g]                  13 [1]
    5  mult [d b] [h]                  13 [2]
    6  mult [e a] [i]                  13 [3]
    7  eqn [g c] []                    57 [1 4]
    8  le [h e] []                     58 [2 5]
    9  mult [h a] [j]                  51 [8 6]
   10  mult [d g] [k]                  69 [5 9 4]
   11  mult [d c] [l]                  sr1 [10 7]
   12  mult [c d] [m]                  13 [11]
 eop

Theorem  77  [11 62 54]                                    0

lt [a b] []
lt [c d] []
mult [b d] [e]
eqn [a e] []
--------------
false

proof

    1  lt [a b] []
    2  lt [c d] []
    3  mult [b d] [e]
    4  eqn [a e] []
    5  le [e a] []                     11 [4]
    6  lt [a e] []                     62 [1 2 3]
    7  false                           54 [6 5]
 eop

Theorem  78  [11 64 54]                                    0

add [a b] [c]
lt [c d] []
add [b d] [e]
eqn [a e] []
-------------
false

proof

    1  add [a b] [c]
    2  lt [c d] []
    3  add [b d] [e]
    4  eqn [a e] []
    5  le [e a] []                     11 [4]
    6  lt [a e] []                     64 [1 2 3]
    7  false                           54 [6 5]
 eop
```

```
Axiom  79                                                     1

add [a b] [c]
add [c d] [e]
add [b d] [f]
add [a f] [g]
-------------
eqn [g e] []

Theorem  80  [79 14]                                          0

add [a b] [c]
add [c d] [e]
add [b d] [f]
add [a f] [g]
-------------
le [g e] []

proof

    1  add [a b] [c]
    2  add [c d] [e]
    3  add [b d] [f]
    4  add [a f] [g]
    5  eqn [g e] []                  79 [1 2 3 4]
    6  le [g e] []                   14 [5]
 eop

Axiom  81                                                     1

mult [a b] [c]
mult [c d] [e]
mult [b d] [f]
mult [a f] [g]
--------------
eqn [g e] []

Theorem  82  [81 14]                                          0

mult [a b] [c]
mult [c d] [e]
mult [b d] [f]
mult [a f] [g]
--------------
le [g e] []

proof

    1  mult [a b] [c]
    2  mult [c d] [e]
    3  mult [b d] [f]
    4  mult [a f] [g]
    5  eqn [g e] []                  81 [1 2 3 4]
    6  le [g e] []                   14 [5]
 eop

Theorem  83  [11 70 54]                                       3

le [a b] []
lt [c d] []
add [b d] [e]
eqn [a e] []
-------------
false

proof

    1  le [a b] []
    2  lt [c d] []
    3  add [b d] [e]
    4  eqn [a e] []
```

```
    5  le [e a] []                     11 [4]
    6  lt [a e] []                     70 [1 2 3]
    7  false                           54 [6 5]
 eop

Axiom  84                                                11

lt [a b] []
add [a c] [d]
add [b c] [e]
-------------
lt [d e] []

Theorem  85  [84 15]                                      0

lt [a b] []
add [a c] [d]
add [b c] [e]
-------------
le [d e] []

proof

    1  lt [a b] []
    2  add [a c] [d]
    3  add [b c] [e]
    4  lt [d e] []                     84 [1 2 3]
    5  le [d e] []                     15 [4]
 eop

Theorem  86  [15 88]                                      0

lt [a b] []
mult [a c] [d]
mult [b c] [e]
--------------
le [d e] []

proof

    1  lt [a b] []
    2  mult [a c] [d]
    3  mult [b c] [e]
    4  le [a b] []                     15 [1]
    5  le [d e] []                     88 [4 2 3]
 eop

UD  87                                                    0

le [a b] []
add [a c] [d]
add [b c] [e]
-------------
le [d e] []

Axiom  88                                                 1

le [a b] []
mult [a c] [d]
mult [b c] [e]
--------------
le [d e] []

UD  89                                                    0

lt [a b] []
lt [0 c] []
add [a c] [d]
mult [b c] [e]
--------------
le [d e] []
```

```
Axiom  90                                                   4

lt [a b] []
lt [0 c] []
mult [a c] [d]
mult [b c] [e]
--------------
lt [d e] []

Theorem  91  [11 84 54]                                     3

add [a b] [c]
add [d b] [e]
eqn [e c] []
lt [d a] []
-------------
false

proof

    1  add [a b] [c]
    2  add [d b] [e]
    3  eqn [e c] []
    4  lt [d a] []
    5  le [c e] []                    11 [3]
    6  lt [e c] []                    84 [4 2 1]
    7  false                          54 [6 5]
 eop

Theorem  92  [1 10 91 59]                                   2

add [a b] [c]
add [d b] [e]
eqn [e c] []
-------------
eqn [d a] []

proof

    1  add [a b] [c]
    2  add [d b] [e]
    3  eqn [e c] []
    4  typen [a] []                   iot [1]
    5  typen [d] []                   iot [2]
    6  trich [a d] []                 1 [4 5]
    7  trich [d a] []                 1 [5 4]
    8  eqn [c e] []                   10 [3]
    9  le [a d] []                    91 [1 2 3 6] dcr2 [6]
   10  le [d a] []                    91 [2 1 8 7] dcr2 [7]
   11  eqn [d a] []                   59 [10 9]
 eop

Theorem  93  [92 14]                                        1

add [a b] [c]
add [d b] [e]
eqn [e c] []
-------------
le [d a] []

proof

    1  add [a b] [c]
    2  add [d b] [e]
    3  eqn [e c] []
    4  eqn [d a] []                   92 [1 2 3]
    5  le [d a] []                    14 [4]
 eop

Axiom  94                                                   2
```

```
add [a b] [c]
add [d b] [e]
lt [e c] []
-------------
lt [d a] []

Theorem  95  [94 15]                                          1

add [a b] [c]
add [d b] [e]
lt [e c] []
-------------
le [d a] []

proof

    1  add [a b] [c]
    2  add [d b] [e]
    3  lt [e c] []
    4  lt [d a] []                     94 [1 2 3]
    5  le [d a] []                     15 [4]
 eop

Theorem  96  [12 16 50]                                       0

add [a b] [c]
add [d b] [e]
add [e c] [f]
-------------
add [d a] [j]

proof

    1  add [a b] [c]
    2  add [d b] [e]
    3  add [e c] [f]
    4  add [c e] [g]                   12 [3]
    5  le [a c] []                     16 [1]
    6  le [d e] []                     16 [2]
    7  add [a e] [h]                   50 [5 4]
    8  add [e a] [i]                   12 [7]
    9  add [d a] [j]                   50 [6 8]
 eop

Theorem  97  [13 16 51]                                       0

add [a b] [c]
add [d b] [e]
mult [e c] [f]
--------------
mult [d a] [j]

proof

    1  add [a b] [c]
    2  add [d b] [e]
    3  mult [e c] [f]
    4  mult [c e] [g]                  13 [3]
    5  le [a c] []                     16 [1]
    6  le [d e] []                     16 [2]
    7  mult [a e] [h]                  51 [5 4]
    8  mult [e a] [i]                  13 [7]
    9  mult [d a] [j]                  51 [6 8]
 eop

Theorem  98  [95 93]                                          0

add [a b] [c]
add [d b] [e]
le [e c] []
```

```
-------------
le [d a] []

proof

    1  add [a b] [c]
    2  add [d b] [e]
    3  le [e c] []
    4  le [d a] []                    95 [1 2 3] 93 [1 2 3]
 eop

Axiom  99                                                2

mult [a b] [c]
mult [d b] [e]
lt [e c] []
--------------
lt [d a] []

Theorem  100  [99 15]                                    1

mult [a b] [c]
mult [d b] [e]
lt [e c] []
--------------
le [d a] []

proof

    1  mult [a b] [c]
    2  mult [d b] [e]
    3  lt [e c] []
    4  lt [d a] []                    99 [1 2 3]
    5  le [d a] []                    15 [4]
 eop

UD  101                                                  0

le [a b] []
mult [c b] [d]
lt [d e] []
--------------
add [c a] [f]

Theorem  102  [11 90 54]                                 3

lt [0 a] []
mult [b a] [c]
mult [d a] [e]
eqn [e c] []
lt [d b] []
--------------
false

proof

    1  lt [0 a] []
    2  mult [b a] [c]
    3  mult [d a] [e]
    4  eqn [e c] []
    5  lt [d b] []
    6  le [c e] []                    11 [4]
    7  lt [e c] []                    90 [5 1 3 2]
    8  false                          54 [7 6]
 eop

Theorem  103  [3 13 16 70 44 83 12 43]                   0

lt [0 a] []
add [b a] [c]
add [d a] [e]
```

```
mult [e c] [f]
--------------
add [d b] [j]

proof

    1  lt [0 a] []
    2  add [b a] [c]
    3  add [d a] [e]
--------------
    4  mult [e c] [f]
    5  typen [b] []                iot [2]
    6  typen [d] []                iot [3]
    7  le [b b] []                 3 [5]
    8  le [d d] []                 3 [6]
    9  mult [c e] [g]              13 [4]
   10  le [b c] []                 16 [2]
   11  lt [d e] []                 70 [8 1 3]
   12  add [b e] [h]               44 [10 9] 83 [7 1 2 10]
   13  add [e b] [i]               12 [12]
   14  add [d b] [j]               43 [11 13]
 eop

UD  104                                                 0

lt [0 a] []
add [b a] [c]
mult [d a] [e]
lt [e c] []
--------------
le [d b] []

Theorem  105  [3 12 16 50 72]                           1

lt [0 a] []
add [b a] [c]
mult [d a] [e]
add [e c] [f]
--------------
add [d b] [j]

proof

    1  lt [0 a] []
    2  add [b a] [c]
    3  mult [d a] [e]
    4  add [e c] [f]
    5  typen [d] []                iot [3]
    6  le [d d] []                 3 [5]
    7  add [c e] [g]               12 [4]
    8  le [b c] []                 16 [2]
    9  add [b e] [h]               50 [8 7]
   10  le [d e] []                 72 [6 1 3]
   11  add [e b] [i]               12 [9]
   12  add [d b] [j]               50 [10 11]
 eop

Theorem  106  [3 13 16 72 44 83 12 50]                  1

lt [0 a] []
add [b a] [c]
mult [d a] [e]
mult [e c] [f]
--------------
add [d b] [j]

proof

    1  lt [0 a] []
    2  add [b a] [c]
    3  mult [d a] [e]
```

```
    4  mult [e c] [f]
    5  typen [b] []                    iot [2]
    6  typen [d] []                    iot [3]
    7  le [b b] []                     3 [5]
    8  le [d d] []                     3 [6]
    9  mult [c e] [g]                  13 [4]
   10  le [b c] []                     16 [2]
   11  le [d e] []                     72 [8 1 3]
   12  add [b e] [h]                   44 [10 9] 83 [7 1 2 10]
   13  add [e b] [i]                   12 [12]
   14  add [d b] [j]                   50 [11 13]
 eop

Theorem  107  [3 13 16 51 72]                              1

lt [0 a] []
add [b a] [c]
mult [d a] [e]
mult [e c] [f]
--------------
mult [d b] [j]

proof

    1  lt [0 a] []
    2  add [b a] [c]
    3  mult [d a] [e]
    4  mult [e c] [f]
    5  typen [d] []                    iot [3]
    6  le [d d] []                     3 [5]
    7  mult [c e] [g]                  13 [4]
    8  le [b c] []                     16 [2]
    9  mult [b e] [h]                  51 [8 7]
   10  le [d e] []                     72 [6 1 3]
   11  mult [e b] [i]                  13 [9]
   12  mult [d b] [j]                  51 [10 11]
 eop

Theorem  108  [12 105]                                     0

lt [0 a] []
mult [b a] [c]
add [d a] [e]
add [e c] [f]
--------------
add [d b] [i]

proof

    1  lt [0 a] []
    2  mult [b a] [c]
    3  add [d a] [e]
    4  add [e c] [f]
    5  add [c e] [g]                   12 [4]
    6  add [b d] [h]                   105 [1 3 2 5]
    7  add [d b] [i]                   12 [6]
 eop

Theorem  109  [13 106 12]                                  0

lt [0 a] []
mult [b a] [c]
add [d a] [e]
mult [e c] [f]
--------------
add [d b] [i]

proof

    1  lt [0 a] []
    2  mult [b a] [c]
```

```
    3  add [d a] [e]
    4  mult [e c] [f]
    5  mult [c e] [g]                  13 [4]
    6  add [b d] [h]                   106 [1 3 2 5]
    7  add [d b] [i]                   12 [6]
 eop

Theorem  110  [13 107]                                       0

lt [0 a] []
mult [b a] [c]
add [d a] [e]
mult [e c] [f]
--------------
mult [d b] [i]

proof

    1  lt [0 a] []
    2  mult [b a] [c]
    3  add [d a] [e]
    4  mult [e c] [f]
    5  mult [c e] [g]                  13 [4]
    6  mult [b d] [h]                  107 [1 3 2 5]
    7  mult [d b] [i]                  13 [6]
 eop

Theorem  111  [1 10 102 59]                                  2

lt [0 a] []
mult [b a] [c]
mult [d a] [e]
eqn [e c] []
--------------
eqn [d b] []

proof

    1  lt [0 a] []
    2  mult [b a] [c]
    3  mult [d a] [e]
    4  eqn [e c] []
    5  typen [b] []                    iot [2]
    6  typen [d] []                    iot [3]
    7  trich [b d] []                  1 [5 6]
    8  trich [d b] []                  1 [6 5]
    9  eqn [c e] []                    10 [4]
   10  le [b d] []                     102 [1 2 3 4 7] dcr2 [7]
   11  le [d b] []                     102 [1 3 2 9 8] dcr2 [8]
   12  eqn [d b] []                    59 [11 10]
 eop

Theorem  112  [111 14]                                       1

lt [0 a] []
mult [b a] [c]
mult [d a] [e]
eqn [e c] []
--------------
le [d b] []

proof

    1  lt [0 a] []
    2  mult [b a] [c]
    3  mult [d a] [e]
    4  eqn [e c] []
    5  eqn [d b] []                    111 [1 2 3 4]
    6  le [d b] []                     14 [5]
 eop
```

```
UD  113                                                         0

lt [0 a] []
mult [b a] [c]
mult [d a] [e]
add [e c] [f]
--------------
add [d b] [g]

UD  114                                                         0

lt [0 a] []
mult [b a] [c]
mult [d a] [e]
mult [e c] [f]
mult [d b] [g]

Theorem  115  [100 112]                                         0

lt [0 a] []
mult [b a] [c]
mult [d a] [e]
le [e c] []
--------------
le [d b] []

proof

    1  lt [0 a] []
    2  mult [b a] [c]
    3  mult [d a] [e]
    4  le [e c] []
    5  le [d b] []                     100 [2 3 4] 112 [1 2 3 4]
 eop




  number of axioms =          28
  number of theorems (provisional) =           0
  number of theorems =          79
  number of underivable ieps =           8
```